\documentstyle[aps,prb,psfig,floats]{revtex}
\begin{document}
\title{Van der Waals clusters in the ultra-quantum limit: a Monte Carlo study}

\author{M. Meierovich, A. Mushinski and M.P. Nightingale }
\address{
  Department of Physics,\\
  University of Rhode Island,\\
  Kingston, RI 02881.
}
\maketitle

\begin{abstract}
Bosonic van der Waals clusters of sizes three, four and five are
studied by diffusion quantum Monte-Carlo techniques. In particular we
study the unbinding transition, the ultra-quantum limit where the
ground state ceases to exist as a bound state.  We discuss the quality
of trial wave functions used in the calculations, the critical behavior
in the vicinity of the unbinding transition, and simple improvements of
the diffusion Monte Carlo algorithm.
\end{abstract}

\section{Introduction}
\label{Introduction_section}

In a previous paper \cite{MushNigh.94}, a form of variational trial
wave function was developed and applied to atomic, few-body systems
consisting of five or less atoms of Ar and Ne interacting via a
Lennard-Jones potential.  In addition, we tested the trial functions
for a hypothetical, light atom resembling Ne but with only half its
mass.  We did not study atoms such as $^{4}$He with larger de Boer
parameters, i.e., systems in which the zero point energy plays a more
important role relative to the potential energy.  Studying such systems
is the main purpose of the present paper.  In fact, we study clusters
in the ultra-quantum unbinding limit, in which the zero-point energy
destroys the bound ground state.  Simple arguments applied to this
unbinding transition predict the way in which the energy vanishes as
the de Boer parameter approaches its critical value, and the nature of
the divergence of the size of the clusters.  Our numerical results are
in agreement with these predictions.

As the de Boer parameter increases, the quality of the wave functions
decreases and, whereas in Ref.~\onlinecite{MushNigh.94} there was no
need to go beyond variational Monte Carlo, we rely in this paper on
diffusion Monte Carlo to improve the variational estimates.  Since the
diffusion Monte Carlo algorithm is based on a short-time expansion of
the imaginary-time evolution operator $\exp(-\tau {\cal H})$, where
$\cal H$ is the Hamiltonian, the algorithm is exact only in the limit
$\tau \to 0$.  In practice, since the computations are done at finite
$\tau$, and the computer time required to obtain a given statistical
accuracy increases as $\tau^{-1}$, it pays to design the diffusion
Monte Carlo algorithm to have a small time-step error. For this purpose
we use a simplified version of the improved diffusion Monte Carlo
algorithm introduced by Umrigar {\it et al.}\cite{UmriNighRu.93}  We
use the trial wave functions of Ref.~\onlinecite{MushNigh.94} to obtain
both variational and diffusion Monte Carlo estimates for the ground
state energy and other expectation values.

The layout of this paper is as follows.  In
Section~\ref{Methods.section} we review diffusion Monte Carlo and the
modifications we made to the algorithm given in
Ref.~\onlinecite{UmriNighRu.93} to make it applicable to Lennard-Jones
bosons.  Trial functions are optimized by minimizing the variance of
the local energy.  In the immediate vicinity of the unbinding
transition this method becomes unstable, a problem that can be solved
straightforwardly, as discussed in Section~\ref{Methods.section}.  We
found that graphical methods were helpful to identify configuration
space regions that dominate the variance of the local energy.  Some of
the details are discussed also in Section~\ref{Methods.section}.
However, this part of our presentation is incomplete in the sense that
our techniques rely in part on color graphics, which are not included
in this paper.  For additional details we refer to
Ref.~\onlinecite{Marks.internet.visualization}.  Section~\ref{Result}
contains numerical results:  quantitative evaluation of the
improvements obtained by the modified diffusion Monte Carlo algorithm;
estimates of ground state energies; and numerical corroboration of the
critical behavior of the unbinding transition, as predicted below [{\it
cf.} Eqs.~(\ref{eq.critical})].

\section{Methods}
\label{Methods.section}

We consider a clusters of $N$ bosonic Lennard-Jones atoms.  The
Lennard-Jones pair potential $v(r)=4\epsilon[ (r/\sigma)^{-12}-
(r/\sigma)^{-6a}]$ in reduced units takes the form $v(r)=r^{-12}-2
r^{-6}$, and the only independent parameter in the Schr\"odinger
equation is the reduced inverse mass $m^{-1}$, a quantity proportional
to the square of the de Boer parameter, $h/\sigma \sqrt{m \epsilon}$, a
dimensionless quantity measuring the importance of quantum mechanical
effects.

A cluster configuration is given by the cartesian coordinates of the
atoms, which form a $3N$-dimensional vector ${\bf R} = ({\bf
r}_1,\dots,{\bf r}_N)$, where ${\bf r}_i$ is a $3$-dimensional vector
specifying the coordinates of atom $i$.  The total potential energy of
the cluster is denoted by ${\cal V({\bf R})}$.

We briefly review the diffusion Monte Carlo algorithm, implemented, as
usual, with importance sampling for which an optimized trial function
${\psi_{\rm T}}({\bf R})$ is introduced.  The variational energy of
this state is written as $E_{\rm T}$. For a given trial wave function
${\psi_{\rm T}}({\bf R})$, one introduces a distribution $f({\bf R},t)
= {\psi_{\rm T}}({\bf R})\psi({\bf R},t)$. Since $\psi({\bf R},t)$
satisfies the Schr\"odinger equation in imaginary time, $f({\bf R},t)$
can be shown \cite{Moskowitz82,Reynolds82} to be a solution of the
equation
\begin{equation}
\label{Schrod}
-{1 \over 2m}{\bf \nabla}^{2}f({\bf R},t) 
+ {1 \over m}{\bf \nabla} \cdot [{\bf V}({\bf R})f({\bf R},t)] - 
S({\bf R})f({\bf R},t) 
= - {\partial f({\bf R},t) \over \partial t}, 
\end{equation}
where the {\em velocity} ${\bf V}$, usually called the {\em quantum
force}, is given by
\begin{equation}
\label{Velocity}
{\bf V}({\bf R}) = ({\bf v}_1,\dots,{\bf v}_N) = 
{{\bf \nabla}{\psi_{\rm T}}({\bf R}) \over {\psi_{\rm T}}({\bf R})} = 
{(\bf{\partial}_1,\dots,{\partial}_N){\psi_{\rm T}}({\bf R}) 
\over {\psi_{\rm T}}({\bf R})},
\end{equation}
and the coefficient of the source term is defined as
\begin{equation}
\label{Source_term}
S({\bf R}) = E_{\rm T} - {\cal E}({\bf R}),
\end{equation}
which in turn is defined in terms of the {\em local energy}
\begin{equation}
\label{Local_energy}
{\cal E}({\bf R}) = {{\cal H}{\psi_{\rm T}}({\bf R}) 
\over {\psi_{\rm T}}({\bf R})} = 
-{1 \over 2m}{{\bf \nabla}^2{\psi_{\rm T}}({\bf R}) 
\over {\psi_{\rm T}}({\bf R})} + {\cal V}({\bf R}).
\end{equation}
Note that the energy in the Schr\"odinger equation was shifted so that
$S$ vanishes if the trial state is an exact energy eigenstate.

The simplest implementation of the diffusion Monte Carlo algorithm is
based on a short-time approximation of $\tilde{G}({\bf R'},{\bf
R},\tau) \equiv \langle {\bf R}|\exp(-\tau {\cal H})|{\bf
R'}\rangle$.  This takes the form of a product of the Green
functions of each of the three operators on the left-hand side of
Eq.~(\ref{Schrod}):
\begin{equation}
\label{Green}
\tilde{G}({\bf R'},{\bf R},\tau m) \approx 
\left(2\pi \tau \right)^{-{3N \over 2}}
\int d{{\bf R''}} 
e^{-{{({{\bf R'}}-{\bf R''})}^2 \over 2 \tau }}
\delta[{\bf R''} - {\bf R} - {\bf V}({\bf R}) \tau] 
e^{-{1 \over 2}[{\cal E}({\bf R'})+{\cal E}({\bf R})]\tau m}.
\end{equation}
The Monte Carlo incarnation of the above expression consists of a
deterministic drift of an initial configuration ${\bf R}$ to a new
configuration $\bf R''$ followed by a random diffusion to a
final configuration $\bf R'$.  Finally, there is a reweighting
based on the initial and final configurations.  According to
Eq.~(\ref{Green}), both drift and diffusion modify the coordinates of
{\it each} atom in the configuration simultaneously.  Alternatively,
one can break up the operators on the left-hand side of
Eq.~(\ref{Schrod}) into single particle operators.  Correspondingly,
one can write a short-time Green function as a product of factors
associated with drift and diffusion of individual atoms, e.g., in the
order defined by the numbering of the atoms $i=1,\dots,N$.  In our
computations we used this alternative approximation, while we kept the
same exponential growth-decay factor as in Eq.~(\ref{Green}), rather
than including a reweighting factor for each single-atom update.

The imaginary-time evolution operator does not uniquely define a
short-time expansion.  The corresponding freedom can be exploited to
extend the range in $\tau$ over which the approximate short-time Green
function agrees with the exact expression to some given accuracy
\cite{UmriNighRu.93}.  In other words, the time-step error of the
diffusion Monte Carlo algorithm can be reduced by adapting the
algorithm so that it can deal more accurately with singular regions of
configuration space.  In this way, one can make simple algorithmic
changes of essentially zero computational cost to improve the
efficiency of the algorithm dramatically.  Indeed, this was the guiding
principle in the design of the diffusion Monte Carlo algorithm
described in Ref.~\onlinecite{UmriNighRu.93}.

When one is dealing with atoms or molecules in which the ``elementary''
particles are the electrons and nuclei, there are numerous sources of
singular behavior:  electron-electron and electron-nucleus collisions
and nodes in the trial function.  In the bosonic case, the ground state
has no nodes and the only singularities are due to inter-atomic
collisions.  For example, the exact Green function of the second term,
i.e., the drift term in Eq.~(\ref{Schrod}) is $\delta[{\bf R''}
- {\bf R}(\tau)]$, where ${\bf R}(t)$ is the position at time $t$
obtained by exact integration of the velocity ${\bf V}$ subject to the
initial condition that ${\bf R}(t)={\bf R}$ at $t=0$.  This exact
expression reduces to the approximation $\delta[{\bf R''} -
{\bf R} - {\bf V}({\bf R})\tau]$ if ${\bf V}$ is assumed constant
during the time $\tau$ in Eq.~(\ref{Green}), but this assumption fails
when two atoms collide, as can be seen as follows.

One of the boundary conditions one would like to impose on the trial
wave function is that the local energy remain finite when the distance
between two atoms vanishes, but unfortunately, we have not been able to
meet this goal. In particular, suppose that the distance $r$ between
two atoms vanishes while all other distances remain finite and
non-zero.  By choosing a trial wave function that behaves as
$\exp(-\sqrt m/5 r^5)$, one can satisfy the condition that for $r \to
0$ the local energy does not diverge as strongly as the potential
energy \cite{MushNigh.94}, i.e., as $r^{-12}$, but only as $r^{-6}$. In
this case, Eq.~(\ref{Velocity}) implies that the velocity diverges as
$r^{-6}\sqrt m$. This divergence implies that for sufficiently small
$r$ the approximation ${\bf R}(\tau)\approx {\bf R}+{\bf V}({\bf
R})\tau$ becomes a poor one, since it is obtained from the assumption
that the velocity is constant during the time $\tau$.  To improve the
approximation following Ref.~\onlinecite{UmriNighRu.93}, we integrate
the speed as given by the differential equation $v=r^{-6}\sqrt m$ for
the two-particle problem and express the resulting average speed in
terms of the initial speed.  Applied to the drift of atom $i$ this
yields
\begin{equation}
\label{mean_vel.1}
\bar{\bf{v}}_{1i} = 
{{-1 + \sqrt[7]{1+7({1\over m})^{1\over 12}v_i^{7\over
6}\tau}}\over
{({1\over m})^{1\over 12} v_i^{7\over 6}\tau}}{\bf v}_i.
\end{equation}
If ${\bf v}_i \tau$, the one-particle drift for atom $i$, is replaced
by $\bar{\bf v}_{1i} \tau$, the original expression is reproduced for
small velocities, while for large velocities the magnitude of the drift
is reduced to $(7 \tau \sqrt m)^{1/7}$.

The problem of the short-range singularity is most pronounced for light
particles. For heavy particles, on the other hand, the wave function is
strongly peaked close to the classical equilibrium position ${\bf
R}_0$, and in this case the approximation of constant velocity can be
improved too.  Suppose we assume a Gaussian approximation for the wave
function
\begin{equation}
\label{quad.app}
{\psi_{\rm T}}({\bf R}) \propto e^{-A ({\bf R} - {\bf R}_{0})^2},
\end{equation}
where ${\bf R}_0$ represents the classical configuration of minimum energy.
In this approximation, the velocity is always directed towards ${\bf R}_0$
and vanishes at that point.  To compute approximately the drift of atom
$i$ for a trial function of this form, one can express $A$
in terms of the local kinetic energy of particle $i$ and its velocity,
given by Eq.~(\ref{Velocity}), as
\begin{equation}
A = {1 \over 6} [ v_i^2-{\partial_i^2 {\psi_{\rm T}}({\bf R})\over
{\psi_{\rm T}}({\bf R})}] ,
\label{A}
\end{equation}
an expression containing only quantities that have to be computed
anyway.
As in the case of the diverging velocity, one can integrate the
velocity exactly and express the result in terms of an average velocity
\begin{equation}
\label{mean_vel.2}
\bar{\bf{v}}_{2i} = {1-e^{-2A\tau} \over 2A\tau} {\bf v}_i.
\end{equation}

The diffusion Monte Carlo algorithm cannot be expected to be efficient
unless values of $\tau$ are chosen so that the order of magnitude of a
typical drift or diffusion step is comparable to the width of the
region in which the wave function is appreciable.  In the classical,
large $m$ limit, $A \tau \approx m \tau \approx 1$.  This implies that
in practice expression~(\ref{mean_vel.2}) for the mean drift velocity
will not differ dramatically form the one given by the original
expression~(\ref{Velocity}).  In fact, we found that the impact of this
modification of the algorithm on the reduction of the time-step error
and increase of efficiency was insignificant.

We made no attempt to construct a sophisticated scheme to interpolate
between the two approximations as given in Eqs.~(\ref{mean_vel.1}) and
(\ref{mean_vel.2}).  Instead, we simply used the average velocity
$\bar{{\bf v}}_i$ in all our computations:
\begin{equation}
\bar{\bf v}_i = \min(\bar{\bf v}_{1i}, \bar{{\bf v}}_{2i}),
\end{equation}
where value of the function $\min$ is the vectors with the smallest
{\em magnitudes}.

Finally, in order to guarantee that the diffusion Monte Carlo algorithm
produces the exact Green function in the ideal case that the trial
function is the exact ground state we include an accept-reject
step\cite{Reynolds82}.  Once all atoms have drifted and diffused, a new
configuration $\bf R'$ has been generated.  This configuration is
accepted with probability
\begin{equation}
p=\min{({|{\psi_{\rm T}}({\bf R'})|^2 \tilde{G}({\bf R},{\bf R'},\tau)
\over
|{\psi_{\rm T}}({\bf R})|^2 \tilde{G}({\bf R'},{\bf R},\tau)}, \, 1)}.
\end{equation}
If the new configuration $\bf R'$ is rejected, the previous
configuration $\bf R$ is kept.  We note that the accept-reject step
requires for its implementation the introduction of an effective time
step $\tau_{\rm eff}$ in some parts of the algorithm.  For more details
see the electronic structure algorithm in
Ref.~\onlinecite{UmriNighRu.93}, which has to be simplified in obvious
ways to be applicable to the current, atomic system.

For a given amount of computer time, the statistical accuracy of the
diffusion Monte Carlo computations can be increased by using optimized
trial functions.  We used trial wave functions described in
Ref.~\onlinecite{MushNigh.94}.  They were optimized by minimization of
$\chi^2$, the variance of the local energy, but we found that this
optimization procedure was not stable close to the unbinding
transition.  This instability can be understood as follows.

The variance of the local energy cannot be evaluated exactly, or even
numerically exactly, for an arbitrary trial state, since this would
require a $3n$-dimensional integration. Instead, one uses a Monte Carlo
approach in which a few thousand states are sampled from the square of
the trial wave function defined by an initial guess for the parameters
to be optimized.  Then, one changes the parameters and estimates the
variance by reweighting configurations with the appropriate ratio of
the current probability density and the one from which the sample was
drawn originally.  This can be done efficiently as long as the the two
wave functions have sufficient overlap.  Once this condition is no
longer satisfied, one generates a new sample from the current
distribution and iterates until the process converges.

The energy of clusters goes to zero when the de Boer parameter
approaches a critical value, and at the same time these clusters grow
in size.  Under these circumstances, Monte Carlo samples of fixed size
tend to consist of configurations predominantly sampled from the tail
of the wave function.  During the optimization of the wave function,
the local energy tends to become the same, physically incorrect
constant for these configurations, and as a consequence the variance of
the local energy as estimated from a sample of a fixed size can be
reduced artificially by choosing an energy even closer to zero. Of
course, the true variance of the trial wave function increases in this
process, but for a sample of fixed size this goes undetected.  We found
that the solution to this instability is quite simple: rather than
minimizing $\chi^2$ one minimizes $\chi^2/E_{\rm T}^2$.

Computations of the ground state energy by variational or diffusion
Monte Carlo satisfy a zero-variance principle, i.e., in the limit where
the trial function is the exact ground state eigenfunction, the
statistical error vanishes.  As a practical consequence, computations
can be made considerably more efficient by a good choice of trial
functions.  In the design of trial wave functions described in detail
in Ref.~\onlinecite{MushNigh.94}, we followed the same procedure as in
Refs.~\onlinecite{CyrusPRL88,CyrusAthens88}:  the trial functions
satisfy boundary conditions associated with (a) the collision of two
atoms and (b) having one atom go off to infinity.  The most likely
configurations, which involve intermediate distances and require most
of the variational freedom of the trial wave functions, are described
by many-body polynomials\cite{MushNigh.94}.

In the process of improving the quality of wave functions it is
important to know what region of configuration space contributes most
to the variance of the local energy.  For instance, it is useful to
know if the quality of the wave function is limited by poorly satisfied
boundary conditions or whether the quality can be improved by adding
more variational parameters.  Another possibility is that the wave
function has too much variational freedom relative to the sample over
which it is optimized.  This might lead to unphysical peaks in the wave
function, which might only show up in the variance of the local energy
obtained from production runs, which sample a much larger number of
configurations than the number present in the sample used for the
optimization of the trial function.

To help answer such questions, we made density plots of the local
error, the deviation of the local energy from its average. As an
illustration, we discuss the case of five-atom clusters.  In fact, we
used superimposed color density plots of both the wave function and the
local error, which contain more information than can be reproduced by
the grey-scale plots reproduced in this paper.  We refer the reader to
Ref.~\onlinecite{Marks.internet.visualization} for the color graphics.

Obviously, the fact that the ground state wave function depends on
$3N-6$ independent coordinate variables, seriously limits any graphical
approach.  For the five-atom clusters we found the following planar cut
through configuration space informative:  four atoms were fixed at the
vertices of a regular tetrahedron, while the fifth particle was located
in a plane that contains two of these vertices and bisects the edge
connecting the two remaining atoms.

\begin{figure}
\centerline{\psfig{figure=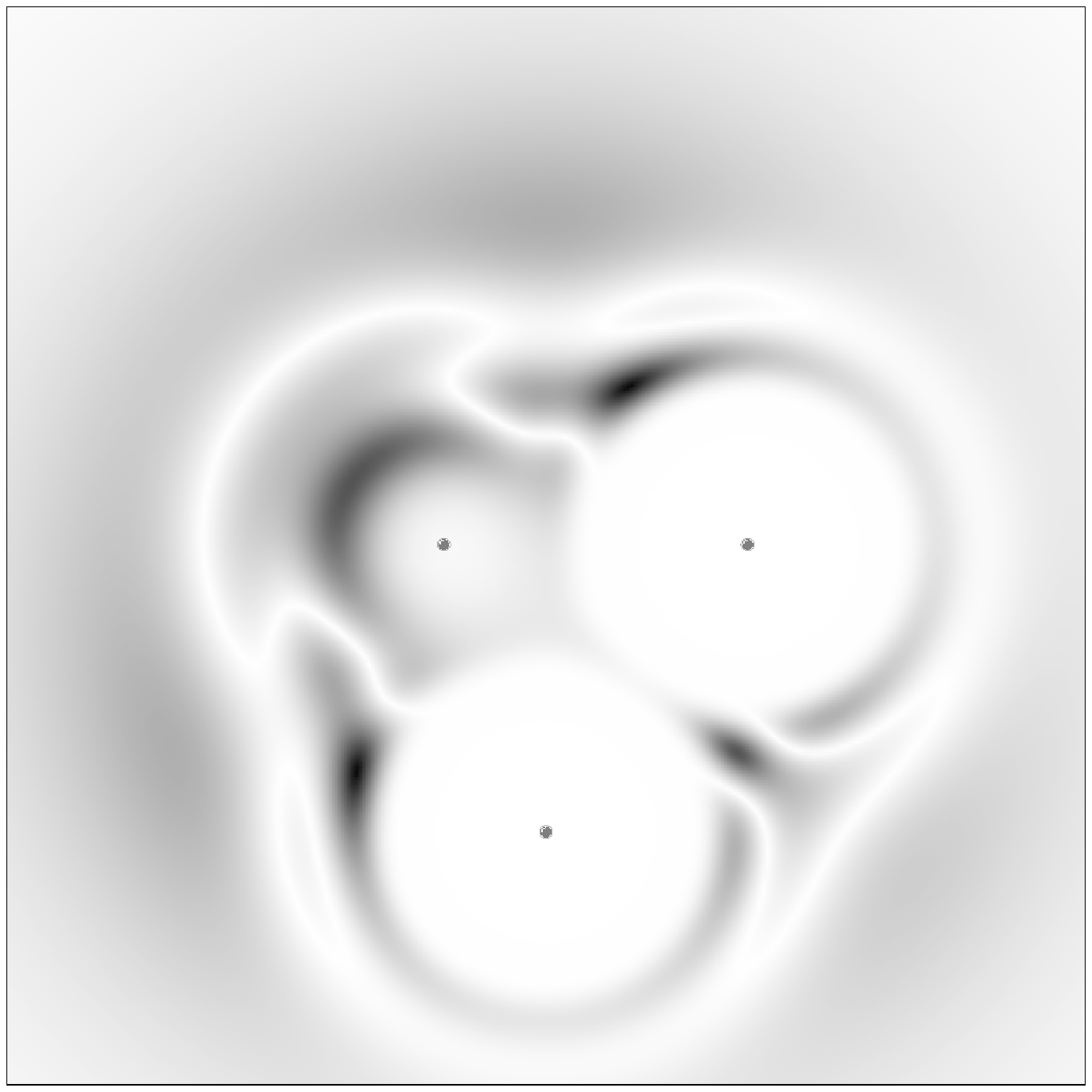}}
\vskip 0.5cm

\caption[Local error for $N=5$, $m^{-1}=0.16$]

{\footnotesize Density plot of the ``local error'' in the geometry
described in the text.  The two dots in the lower right-hand corner are
the two in-plane vertices of the tetrahedron; the one in the upper left
corner is the projection of the two out-of-plane vertices.  The length
of the tetrahedron edges is 1.3 and $m^{-1}=0.16$.  The darker the
region, the more it contributes to $\chi^2$.  Note that the dark region
in the lower right is a cut through the banana-shaped dark region in
the upper left.  The regions of the largest local error are the two
symmetrically located regions where the wandering atom is close to
three others.  White lines are are cuts through nodal surfaces of the
local error and have no physical significance.}

\label{fig.n5.e.16.ps}
\end{figure}

\begin{figure}
\centerline{\psfig{figure=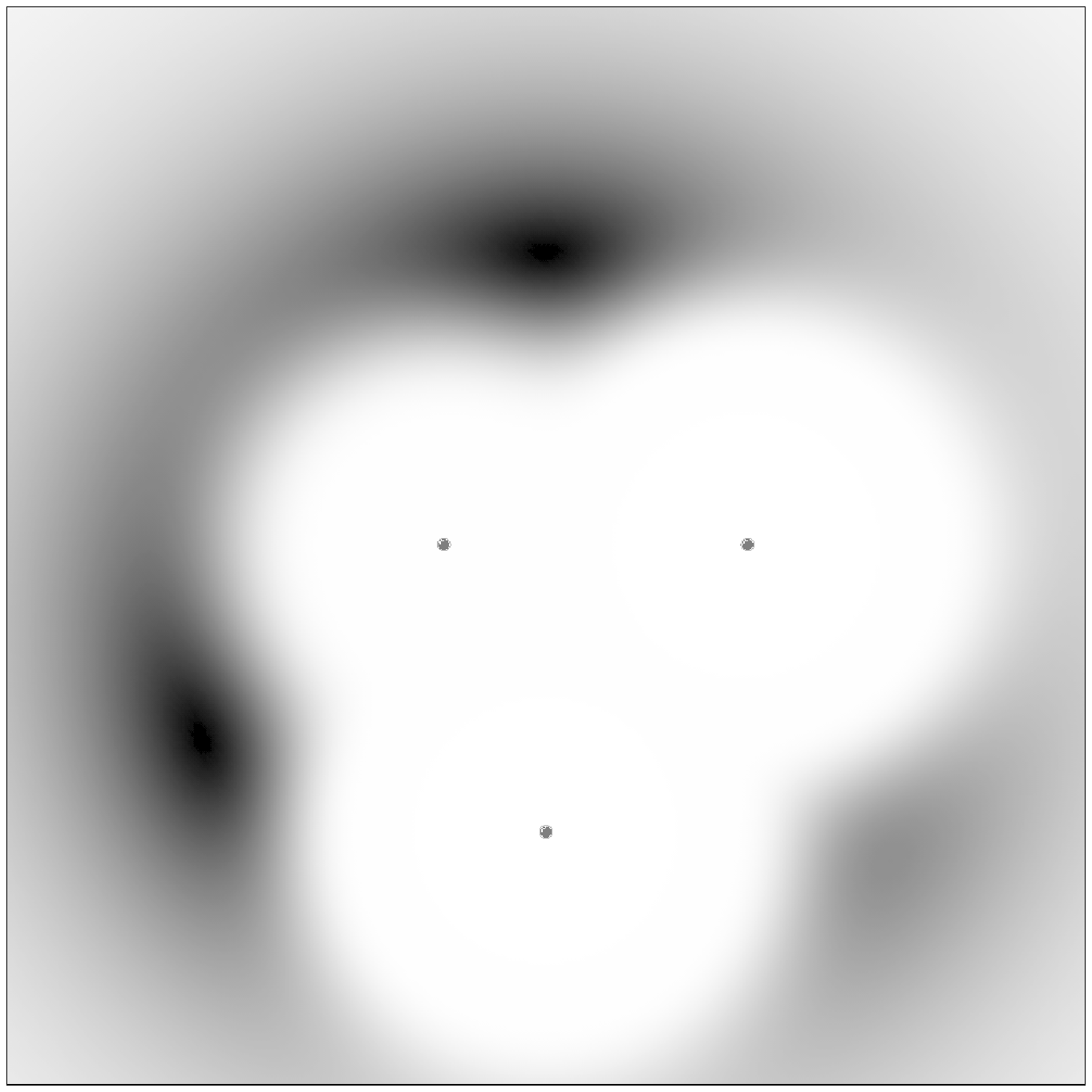}}
\vskip 0.5cm

\caption[$\psi^2$ for $N=5$, $m^{-1}=0.16$]{\footnotesize Density plot
of $|\psi_T|^2$ as a function of the position of the fifth particle in
the geometry described in the text and the figure caption of
Fig.~\ref{fig.n5.e.16.ps}.  Dark regions correspond to high probability
density.  Note that the regions with the largest local error are
contained in the region where the wave function becomes small because
of repulsive potential at short range.}

\label{fig.n5.psi.16.ps}
\end{figure}

Figs. \ref{fig.n5.e.16.ps} and \ref{fig.n5.psi.16.ps} strongly suggest
which regions of configuration space contribute most to $\chi^2$ for
case $N=5$.  For the interpretation of the density plots the following
convention should be used: zero intensity (white) corresponds to a
minimum, while full intensity (black) corresponds to a maximum of the
plotted function.  Fig.~\ref{fig.n5.e.16.ps} represents the density
plot of the weighted ``local error'', defined as $|({\cal E} - E_{\rm
T})\psi_{\rm T}|$, [{\it cf.} Eq.~(\ref{Local_energy})] as a function
of the position of the fifth, wandering atom.  Note that the quantity
$\chi^2$ to be minimized in the optimization of the trial functions is
the configurational integral of the square of this quantity, apart from
a normalization constant.  Fig.~\ref{fig.n5.psi.16.ps} shows the
dependence of $\psi_{\rm T}^2$, on the position of the fifth particle
while the other atoms are fixed in the geometry discussed above.

The conclusion we draw from the density plots is that the trial
function fails particularly in regions where more than two atoms
collide and we see that of these the the local error is largest
whenever four atoms are close.  Unfortunately, so far we have not been
able to find trial functions without this problem, i.e., trial
functions without the $r^{-12}$ divergence of the local energy which
occurs when more than two atoms collide, and without the $r^{-6}$
divergence of two particles in the even distant presence of a third.

\section{Results}
\label{Result}

\subsection{Groundstate energy and time-step error}
\label{Timestep.error}

We present results for the time-step error, discussed in
Section~\ref{Methods.section} and compare the version of the diffusion
Monte Carlo algorithm summarized above with a simple version of the
algorithm in which (a) the velocity is treated as a constant for the
integration of the short-time drift Green function; (b) each move is
unconditionally accepted rather than the result of an accept-reject
step, so that $\tau_{\rm eff}=\tau$.

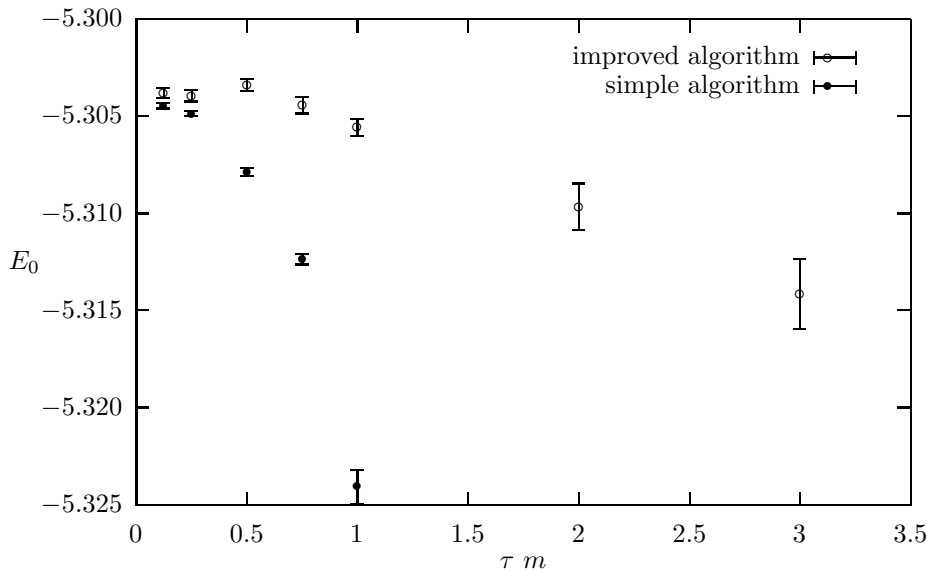
\begin{figure}
\centerline{
\setlength{\unitlength}{0.240900pt}
\ifx\plotpoint\undefined\newsavebox{\plotpoint}\fi
\begin{picture}(1500,900)(0,0)
\font\gnuplot=cmr10 at 10pt
\gnuplot
\sbox{\plotpoint}{\rule[-0.200pt]{0.400pt}{0.400pt}}%
\put(220.0,113.0){\rule[-0.200pt]{0.400pt}{184.048pt}}
\put(220.0,113.0){\rule[-0.200pt]{4.818pt}{0.400pt}}
\put(198,113){\makebox(0,0)[r]{$-5.325$}}
\put(1416.0,113.0){\rule[-0.200pt]{4.818pt}{0.400pt}}
\put(220.0,266.0){\rule[-0.200pt]{4.818pt}{0.400pt}}
\put(198,266){\makebox(0,0)[r]{$-5.320$}}
\put(1416.0,266.0){\rule[-0.200pt]{4.818pt}{0.400pt}}
\put(220.0,419.0){\rule[-0.200pt]{4.818pt}{0.400pt}}
\put(198,419){\makebox(0,0)[r]{$-5.315$}}
\put(1416.0,419.0){\rule[-0.200pt]{4.818pt}{0.400pt}}
\put(220.0,571.0){\rule[-0.200pt]{4.818pt}{0.400pt}}
\put(198,571){\makebox(0,0)[r]{$-5.310$}}
\put(1416.0,571.0){\rule[-0.200pt]{4.818pt}{0.400pt}}
\put(220.0,724.0){\rule[-0.200pt]{4.818pt}{0.400pt}}
\put(198,724){\makebox(0,0)[r]{$-5.305$}}
\put(1416.0,724.0){\rule[-0.200pt]{4.818pt}{0.400pt}}
\put(220.0,877.0){\rule[-0.200pt]{4.818pt}{0.400pt}}
\put(198,877){\makebox(0,0)[r]{$-5.300$}}
\put(1416.0,877.0){\rule[-0.200pt]{4.818pt}{0.400pt}}
\put(220.0,113.0){\rule[-0.200pt]{0.400pt}{4.818pt}}
\put(220,68){\makebox(0,0){0}}
\put(220.0,857.0){\rule[-0.200pt]{0.400pt}{4.818pt}}
\put(394.0,113.0){\rule[-0.200pt]{0.400pt}{4.818pt}}
\put(394,68){\makebox(0,0){0.5}}
\put(394.0,857.0){\rule[-0.200pt]{0.400pt}{4.818pt}}
\put(567.0,113.0){\rule[-0.200pt]{0.400pt}{4.818pt}}
\put(567,68){\makebox(0,0){1}}
\put(567.0,857.0){\rule[-0.200pt]{0.400pt}{4.818pt}}
\put(741.0,113.0){\rule[-0.200pt]{0.400pt}{4.818pt}}
\put(741,68){\makebox(0,0){1.5}}
\put(741.0,857.0){\rule[-0.200pt]{0.400pt}{4.818pt}}
\put(915.0,113.0){\rule[-0.200pt]{0.400pt}{4.818pt}}
\put(915,68){\makebox(0,0){2}}
\put(915.0,857.0){\rule[-0.200pt]{0.400pt}{4.818pt}}
\put(1089.0,113.0){\rule[-0.200pt]{0.400pt}{4.818pt}}
\put(1089,68){\makebox(0,0){2.5}}
\put(1089.0,857.0){\rule[-0.200pt]{0.400pt}{4.818pt}}
\put(1262.0,113.0){\rule[-0.200pt]{0.400pt}{4.818pt}}
\put(1262,68){\makebox(0,0){3}}
\put(1262.0,857.0){\rule[-0.200pt]{0.400pt}{4.818pt}}
\put(1436.0,113.0){\rule[-0.200pt]{0.400pt}{4.818pt}}
\put(1436,68){\makebox(0,0){3.5}}
\put(1436.0,857.0){\rule[-0.200pt]{0.400pt}{4.818pt}}
\put(220.0,113.0){\rule[-0.200pt]{292.934pt}{0.400pt}}
\put(1436.0,113.0){\rule[-0.200pt]{0.400pt}{184.048pt}}
\put(220.0,877.0){\rule[-0.200pt]{292.934pt}{0.400pt}}
\put(45,495){\makebox(0,0){$E_0$}}
\put(828,23){\makebox(0,0){$\tau~m$}}
\put(220.0,113.0){\rule[-0.200pt]{0.400pt}{184.048pt}}
\put(1262,816){\makebox(0,0)[r]{${\rm improved~algorithm}$}}
\put(1306,816){\circle{12}}
\put(1262,445){\circle{12}}
\put(915,581){\circle{12}}
\put(567,707){\circle{12}}
\put(481,741){\circle{12}}
\put(394,773){\circle{12}}
\put(307,756){\circle{12}}
\put(263,760){\circle{12}}
\put(1284.0,816.0){\rule[-0.200pt]{15.899pt}{0.400pt}}
\put(1284.0,806.0){\rule[-0.200pt]{0.400pt}{4.818pt}}
\put(1350.0,806.0){\rule[-0.200pt]{0.400pt}{4.818pt}}
\put(1262.0,390.0){\rule[-0.200pt]{0.400pt}{26.258pt}}
\put(1252.0,390.0){\rule[-0.200pt]{4.818pt}{0.400pt}}
\put(1252.0,499.0){\rule[-0.200pt]{4.818pt}{0.400pt}}
\put(915.0,545.0){\rule[-0.200pt]{0.400pt}{17.586pt}}
\put(905.0,545.0){\rule[-0.200pt]{4.818pt}{0.400pt}}
\put(905.0,618.0){\rule[-0.200pt]{4.818pt}{0.400pt}}
\put(567.0,693.0){\rule[-0.200pt]{0.400pt}{6.504pt}}
\put(557.0,693.0){\rule[-0.200pt]{4.818pt}{0.400pt}}
\put(557.0,720.0){\rule[-0.200pt]{4.818pt}{0.400pt}}
\put(481.0,728.0){\rule[-0.200pt]{0.400pt}{6.263pt}}
\put(471.0,728.0){\rule[-0.200pt]{4.818pt}{0.400pt}}
\put(471.0,754.0){\rule[-0.200pt]{4.818pt}{0.400pt}}
\put(394.0,763.0){\rule[-0.200pt]{0.400pt}{4.577pt}}
\put(384.0,763.0){\rule[-0.200pt]{4.818pt}{0.400pt}}
\put(384.0,782.0){\rule[-0.200pt]{4.818pt}{0.400pt}}
\put(307.0,747.0){\rule[-0.200pt]{0.400pt}{4.336pt}}
\put(297.0,747.0){\rule[-0.200pt]{4.818pt}{0.400pt}}
\put(297.0,765.0){\rule[-0.200pt]{4.818pt}{0.400pt}}
\put(263.0,752.0){\rule[-0.200pt]{0.400pt}{3.854pt}}
\put(253.0,752.0){\rule[-0.200pt]{4.818pt}{0.400pt}}
\put(253.0,768.0){\rule[-0.200pt]{4.818pt}{0.400pt}}
\put(1262,771){\makebox(0,0)[r]{${\rm simple~algorithm}$}}
\put(1306,771){\circle*{12}}
\put(567,142){\circle*{12}}
\put(481,499){\circle*{12}}
\put(394,636){\circle*{12}}
\put(307,728){\circle*{12}}
\put(263,740){\circle*{12}}
\put(1284.0,771.0){\rule[-0.200pt]{15.899pt}{0.400pt}}
\put(1284.0,761.0){\rule[-0.200pt]{0.400pt}{4.818pt}}
\put(1350.0,761.0){\rule[-0.200pt]{0.400pt}{4.818pt}}
\put(567.0,115.0){\rule[-0.200pt]{0.400pt}{12.768pt}}
\put(557.0,115.0){\rule[-0.200pt]{4.818pt}{0.400pt}}
\put(557.0,168.0){\rule[-0.200pt]{4.818pt}{0.400pt}}
\put(481.0,491.0){\rule[-0.200pt]{0.400pt}{3.854pt}}
\put(471.0,491.0){\rule[-0.200pt]{4.818pt}{0.400pt}}
\put(471.0,507.0){\rule[-0.200pt]{4.818pt}{0.400pt}}
\put(394.0,630.0){\rule[-0.200pt]{0.400pt}{3.132pt}}
\put(384.0,630.0){\rule[-0.200pt]{4.818pt}{0.400pt}}
\put(384.0,643.0){\rule[-0.200pt]{4.818pt}{0.400pt}}
\put(307.0,724.0){\rule[-0.200pt]{0.400pt}{1.927pt}}
\put(297.0,724.0){\rule[-0.200pt]{4.818pt}{0.400pt}}
\put(297.0,732.0){\rule[-0.200pt]{4.818pt}{0.400pt}}
\put(263.0,736.0){\rule[-0.200pt]{0.400pt}{2.168pt}}
\put(253.0,736.0){\rule[-0.200pt]{4.818pt}{0.400pt}}
\put(253.0,745.0){\rule[-0.200pt]{4.818pt}{0.400pt}}
\end{picture}} 
\vskip 0.5cm
\caption[f1]{Estimate of the ground state energy $E_0$ as 
a function of time step for Ne$_{5}$}
\label{fig.Ne5}
\end{figure}

We compared the simple and improved diffusion Monte Carlo algorithms
for clusters of Ar, Ne and hypothetical ``{\tiny ${1\over 2}$}-Ne''
atoms with sizes in the range $N = 3,\,4$ and $5$.  Figs.
\ref{fig.Ne5}, and \ref{fig.hNe5} display plots obtained for estimates
of the ground state energy $E_0$ for Ne and ``{\tiny ${1\over 2}$}-Ne''
clusters with $N = 5$; the behavior for the smaller systems is
analogous.  As expected, the reduction of the time-step error is
greatest for the lighter atoms.  For Ar no significant improvement of
the algorithm was achieved, and only an approximate reversal of the
sign of the time-step error occurred.

Table~\ref{tab.compare} displays estimates of the ground state energy
obtained by variational Monte Carlo\cite{MushNigh.94} and the improved
diffusion Monte Carlo algorithm.  In addition, the table contains
information pertaining to the magnitude of the bias of the variational
estimate due to the fact that $E_{\rm T}$ is an upper bound of the true
ground state energy.  The tightness of this bound is determined by the
quality of the trial wave function.  More in detail, the variance of
the local energy is defined by
\begin{equation}
\chi^2={\langle \psi_{\rm T}|({\cal H}-E_0)^2|\psi_{\rm T} \rangle
\over \langle \psi_{\rm T}|\psi_{\rm T} \rangle},
\end{equation}
and the following inequality holds (see Ref.~\onlinecite{MushNigh.94}
for details and references):
\begin{equation}
0<E_{\rm T}-E_0<{\chi^2\over E_1-E_0},
\label{eq.bound}
\end{equation}
where $E_1$ is the energy of the first, totally symmetric excited
state.  To estimate the number of correct digits in the variational
estimate of the ground state
we use the following quantity:
\begin{equation}
Q'  = -\log_{10} \frac{\chi^2}{(E_1 - E_{\rm T})|E_{\rm T}|}.
\label{eq.Qp}
\end{equation}
It is also of interest to know how tight a bound the right-hand side of
inequality (\ref{eq.bound}) provides.  This is measured by the
following ratio
\begin{equation}
R = \frac{\chi^2}{(E_1 - E_{\rm T})(E_{\rm T} - E_0)}.
\end{equation}
The results are shown in Table~\ref{tab.compare}.  Quite remarkably,
the bound given in Eq.~(\ref{eq.bound}) is very tight.

\begin{figure}
\centerline{
\setlength{\unitlength}{0.240900pt}
\ifx\plotpoint\undefined\newsavebox{\plotpoint}\fi
\begin{picture}(1500,900)(0,0)
\font\gnuplot=cmr10 at 10pt
\gnuplot
\sbox{\plotpoint}{\rule[-0.200pt]{0.400pt}{0.400pt}}%
\put(220.0,113.0){\rule[-0.200pt]{0.400pt}{184.048pt}}
\put(220.0,113.0){\rule[-0.200pt]{4.818pt}{0.400pt}}
\put(198,113){\makebox(0,0)[r]{$-4.120$}}
\put(1416.0,113.0){\rule[-0.200pt]{4.818pt}{0.400pt}}
\put(220.0,189.0){\rule[-0.200pt]{4.818pt}{0.400pt}}
\put(198,189){\makebox(0,0)[r]{$-4.115$}}
\put(1416.0,189.0){\rule[-0.200pt]{4.818pt}{0.400pt}}
\put(220.0,266.0){\rule[-0.200pt]{4.818pt}{0.400pt}}
\put(198,266){\makebox(0,0)[r]{$-4.110$}}
\put(1416.0,266.0){\rule[-0.200pt]{4.818pt}{0.400pt}}
\put(220.0,342.0){\rule[-0.200pt]{4.818pt}{0.400pt}}
\put(198,342){\makebox(0,0)[r]{$-4.105$}}
\put(1416.0,342.0){\rule[-0.200pt]{4.818pt}{0.400pt}}
\put(220.0,419.0){\rule[-0.200pt]{4.818pt}{0.400pt}}
\put(198,419){\makebox(0,0)[r]{$-4.100$}}
\put(1416.0,419.0){\rule[-0.200pt]{4.818pt}{0.400pt}}
\put(220.0,495.0){\rule[-0.200pt]{4.818pt}{0.400pt}}
\put(198,495){\makebox(0,0)[r]{$-4.095$}}
\put(1416.0,495.0){\rule[-0.200pt]{4.818pt}{0.400pt}}
\put(220.0,571.0){\rule[-0.200pt]{4.818pt}{0.400pt}}
\put(198,571){\makebox(0,0)[r]{$-4.090$}}
\put(1416.0,571.0){\rule[-0.200pt]{4.818pt}{0.400pt}}
\put(220.0,648.0){\rule[-0.200pt]{4.818pt}{0.400pt}}
\put(198,648){\makebox(0,0)[r]{$-4.085$}}
\put(1416.0,648.0){\rule[-0.200pt]{4.818pt}{0.400pt}}
\put(220.0,724.0){\rule[-0.200pt]{4.818pt}{0.400pt}}
\put(198,724){\makebox(0,0)[r]{$-4.080$}}
\put(1416.0,724.0){\rule[-0.200pt]{4.818pt}{0.400pt}}
\put(220.0,801.0){\rule[-0.200pt]{4.818pt}{0.400pt}}
\put(198,801){\makebox(0,0)[r]{$-4.075$}}
\put(1416.0,801.0){\rule[-0.200pt]{4.818pt}{0.400pt}}
\put(220.0,877.0){\rule[-0.200pt]{4.818pt}{0.400pt}}
\put(198,877){\makebox(0,0)[r]{$-4.070$}}
\put(1416.0,877.0){\rule[-0.200pt]{4.818pt}{0.400pt}}
\put(220.0,113.0){\rule[-0.200pt]{0.400pt}{4.818pt}}
\put(220,68){\makebox(0,0){0}}
\put(220.0,857.0){\rule[-0.200pt]{0.400pt}{4.818pt}}
\put(463.0,113.0){\rule[-0.200pt]{0.400pt}{4.818pt}}
\put(463,68){\makebox(0,0){0.5}}
\put(463.0,857.0){\rule[-0.200pt]{0.400pt}{4.818pt}}
\put(706.0,113.0){\rule[-0.200pt]{0.400pt}{4.818pt}}
\put(706,68){\makebox(0,0){1}}
\put(706.0,857.0){\rule[-0.200pt]{0.400pt}{4.818pt}}
\put(950.0,113.0){\rule[-0.200pt]{0.400pt}{4.818pt}}
\put(950,68){\makebox(0,0){1.5}}
\put(950.0,857.0){\rule[-0.200pt]{0.400pt}{4.818pt}}
\put(1193.0,113.0){\rule[-0.200pt]{0.400pt}{4.818pt}}
\put(1193,68){\makebox(0,0){2}}
\put(1193.0,857.0){\rule[-0.200pt]{0.400pt}{4.818pt}}
\put(1436.0,113.0){\rule[-0.200pt]{0.400pt}{4.818pt}}
\put(1436,68){\makebox(0,0){2.5}}
\put(1436.0,857.0){\rule[-0.200pt]{0.400pt}{4.818pt}}
\put(220.0,113.0){\rule[-0.200pt]{292.934pt}{0.400pt}}
\put(1436.0,113.0){\rule[-0.200pt]{0.400pt}{184.048pt}}
\put(220.0,877.0){\rule[-0.200pt]{292.934pt}{0.400pt}}
\put(45,495){\makebox(0,0){$E_0$}}
\put(828,23){\makebox(0,0){$\tau~m$}}
\put(220.0,113.0){\rule[-0.200pt]{0.400pt}{184.048pt}}
\put(1193,266){\makebox(0,0)[r]{${\rm improved~algorithm}$}}
\put(1237,266){\circle{12}}
\put(1193,625){\circle{12}}
\put(706,768){\circle{12}}
\put(463,793){\circle{12}}
\put(400,803){\circle{12}}
\put(342,795){\circle{12}}
\put(281,792){\circle{12}}
\put(250,791){\circle{12}}
\put(1215.0,266.0){\rule[-0.200pt]{15.899pt}{0.400pt}}
\put(1215.0,256.0){\rule[-0.200pt]{0.400pt}{4.818pt}}
\put(1281.0,256.0){\rule[-0.200pt]{0.400pt}{4.818pt}}
\put(1193.0,594.0){\rule[-0.200pt]{0.400pt}{14.695pt}}
\put(1183.0,594.0){\rule[-0.200pt]{4.818pt}{0.400pt}}
\put(1183.0,655.0){\rule[-0.200pt]{4.818pt}{0.400pt}}
\put(706.0,753.0){\rule[-0.200pt]{0.400pt}{6.986pt}}
\put(696.0,753.0){\rule[-0.200pt]{4.818pt}{0.400pt}}
\put(696.0,782.0){\rule[-0.200pt]{4.818pt}{0.400pt}}
\put(463.0,785.0){\rule[-0.200pt]{0.400pt}{3.613pt}}
\put(453.0,785.0){\rule[-0.200pt]{4.818pt}{0.400pt}}
\put(453.0,800.0){\rule[-0.200pt]{4.818pt}{0.400pt}}
\put(400.0,796.0){\rule[-0.200pt]{0.400pt}{3.373pt}}
\put(390.0,796.0){\rule[-0.200pt]{4.818pt}{0.400pt}}
\put(390.0,810.0){\rule[-0.200pt]{4.818pt}{0.400pt}}
\put(342.0,788.0){\rule[-0.200pt]{0.400pt}{3.373pt}}
\put(332.0,788.0){\rule[-0.200pt]{4.818pt}{0.400pt}}
\put(332.0,802.0){\rule[-0.200pt]{4.818pt}{0.400pt}}
\put(281.0,787.0){\rule[-0.200pt]{0.400pt}{2.650pt}}
\put(271.0,787.0){\rule[-0.200pt]{4.818pt}{0.400pt}}
\put(271.0,798.0){\rule[-0.200pt]{4.818pt}{0.400pt}}
\put(250.0,786.0){\rule[-0.200pt]{0.400pt}{2.650pt}}
\put(240.0,786.0){\rule[-0.200pt]{4.818pt}{0.400pt}}
\put(240.0,797.0){\rule[-0.200pt]{4.818pt}{0.400pt}}
\put(1193,221){\makebox(0,0)[r]{${\rm simple~algorithm}$}}
\put(1237,221){\circle*{12}}
\put(463,332){\circle*{12}}
\put(400,641){\circle*{12}}
\put(342,735){\circle*{12}}
\put(281,790){\circle*{12}}
\put(250,793){\circle*{12}}
\put(1215.0,221.0){\rule[-0.200pt]{15.899pt}{0.400pt}}
\put(1215.0,211.0){\rule[-0.200pt]{0.400pt}{4.818pt}}
\put(1281.0,211.0){\rule[-0.200pt]{0.400pt}{4.818pt}}
\put(463.0,189.0){\rule[-0.200pt]{0.400pt}{68.897pt}}
\put(453.0,189.0){\rule[-0.200pt]{4.818pt}{0.400pt}}
\put(453.0,475.0){\rule[-0.200pt]{4.818pt}{0.400pt}}
\put(400.0,582.0){\rule[-0.200pt]{0.400pt}{28.185pt}}
\put(390.0,582.0){\rule[-0.200pt]{4.818pt}{0.400pt}}
\put(390.0,699.0){\rule[-0.200pt]{4.818pt}{0.400pt}}
\put(342.0,726.0){\rule[-0.200pt]{0.400pt}{4.336pt}}
\put(332.0,726.0){\rule[-0.200pt]{4.818pt}{0.400pt}}
\put(332.0,744.0){\rule[-0.200pt]{4.818pt}{0.400pt}}
\put(281.0,783.0){\rule[-0.200pt]{0.400pt}{3.373pt}}
\put(271.0,783.0){\rule[-0.200pt]{4.818pt}{0.400pt}}
\put(271.0,797.0){\rule[-0.200pt]{4.818pt}{0.400pt}}
\put(250.0,787.0){\rule[-0.200pt]{0.400pt}{2.891pt}}
\put(240.0,787.0){\rule[-0.200pt]{4.818pt}{0.400pt}}
\put(240.0,799.0){\rule[-0.200pt]{4.818pt}{0.400pt}}
\end{picture}} 
\vskip 0.5cm
\caption[f1]{Estimate of the ground state energy $E_0$ as 
a function of time step for {\tiny ${1\over 2}$}-Ne$_5$}
\label{fig.hNe5}
\end{figure}

\begin{table}
\caption[tabcomp]{
\small \parindent .5cm \narrower
Diffusion Monte Carlo estimates of the ground state energies $E_0$ for
noble gases Ar and Ne, and hypothetical ``{\tiny ${1\over 2}$}-Ne'',
compared with variational Monte Carlo estimates taken from
Ref.~\onlinecite{MushNigh.94}.  Standard errors in the last digits are
given in parentheses.
\par\vspace{5mm}}
\begin{tabular} {rcdddd}
\multicolumn{1}{c}{}&
\multicolumn{1}{c}{$N$} &
\multicolumn{1}{c}{$E_{\rm T}$} &
\multicolumn{1}{c}{$E_0$} &
\multicolumn{1}{c}{$Q'$} &
\multicolumn{1}{c}{$R$} \\
\tableline

Ar			&3			&-2.553335364(1)		&-2.553335375(2)	&11.9		&---		\\
Ne			&			&-1.7195589(3)			&-1.7195586(5)		&7.40		&---		\\
{\tiny $1\over 2$}-Ne	&			&-1.308443(2)			&-1.308444(1)		&5.95		&1.5		\\ \hline
Ar			&4			&-5.1182368(2)			&-5.1182376(4)		&7.53		&---		\\
Ne			&			&-3.464174(8)			&-3.464229(13)		&4.67		&1.4		\\
{\tiny $1\over 2$}-Ne	&			&-2.64356(3)			&-2.64383(4)		&3.74		&1.8		\\ \hline
Ar			&5			&-7.78598(1)			&-7.7862(5)		&4.23		&2.1		\\
Ne			&			&-5.29948(8)			&-5.3037(3)		&2.79		&2.0		\\
{\tiny $1\over 2$}-Ne	&			&-4.0669(1)			&-4.0755(5)		&2.55		&2.0		\\
\end{tabular}
 
\label{tab.compare}
\end{table}

\subsection{Unbinding transition}
\label{sec.unbinding}

A severe test for the accuracy of a cluster trial function is its
performance in the strong quantum limit, i.e., for large values of the
de Boer parameter.  In particular, we discuss results in the vicinity
of the unbinding transition, where the cluster ceases to possess a
bound state.

The ground state of the $^4\mbox{He}$ dimer is believed to be a
(weakly) bound state. Since the ground state energy presumably
decreases with cluster size and other systems have smaller de Boer
parameters, we can safely assume that the unbinding transition for
boson clusters is inaccessible experimentally.  However, the transition
does occur at finite cluster size for $^3\mbox{He}$, and it makes sense
to use the boson case as a simpler test case for the trial functions.

A second issue of theoretical interest is the behavior of energy and
cluster size as a function of the de Boer parameter in the vicinity of
the unbinding transition. This transition plays the role of a critical
point and, in fact, has many features in common with a wetting
transition \cite{wetting}.

The following critical behavior is expected for the ground state energy
$E_0$ and the average size $\langle r\rangle$ (as defined below) of the
cluster
\begin{eqnarray}
E_0 & \sim & (\Delta m)^2, \nonumber \\
\langle r \rangle & \sim & (\Delta m)^{-1},
\label{eq.critical}
\end{eqnarray}
for ${m \downarrow m_{\rm c}}$ where $\Delta m = m-m_{\rm c}$ with
$m_{\rm c}$ the critical value of the dimensionless mass.

The critical behavior given in Eq.~(\ref{eq.critical}) can be made
plausible as follows.  For the simple case of a dimer one can show this
directly, and the mathematical mechanism that yields
Eqs.~(\ref{eq.critical}) is the following.  Two scattering states wave
functions forming a complex conjugate pair merge at zero momentum to
produce two states with ``complex momentum": a physically
acceptable bound state and a state with unacceptable behavior at
infinity.  This mechanism is probably not limited to the dimer, and
therefore it is quite plausible that Eqs.~(\ref{eq.critical}) apply in
general to clusters of any finite size.

\begin{figure}
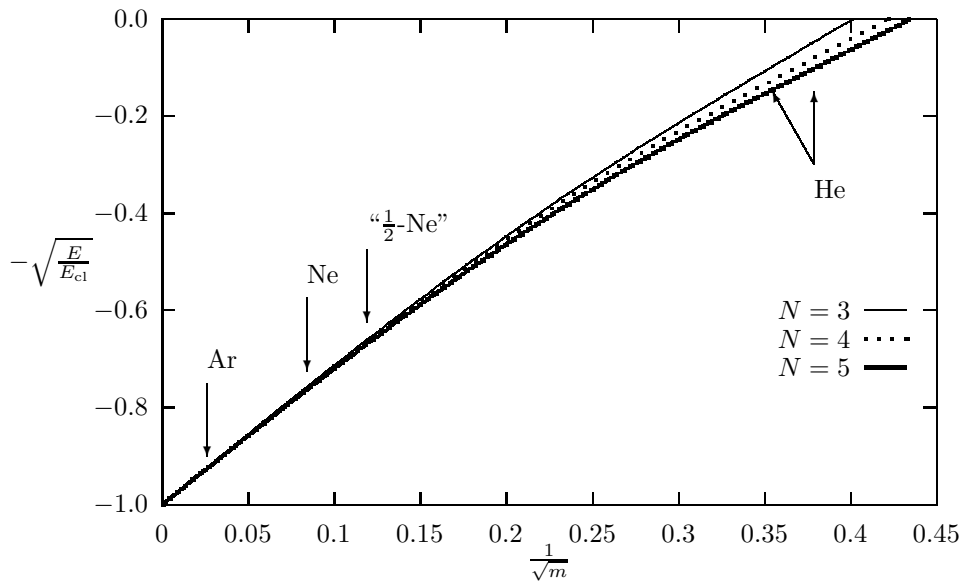

\centerline{\input ./e} 
\vskip 0.5cm
\caption[f1]{Fitting curves for diffusion Monte Carlo estimates of the
ground state energy for clusters of sizes $N = 3,4$ and $5$.}
\label{fig.comb}
\end{figure}

\begin{figure}
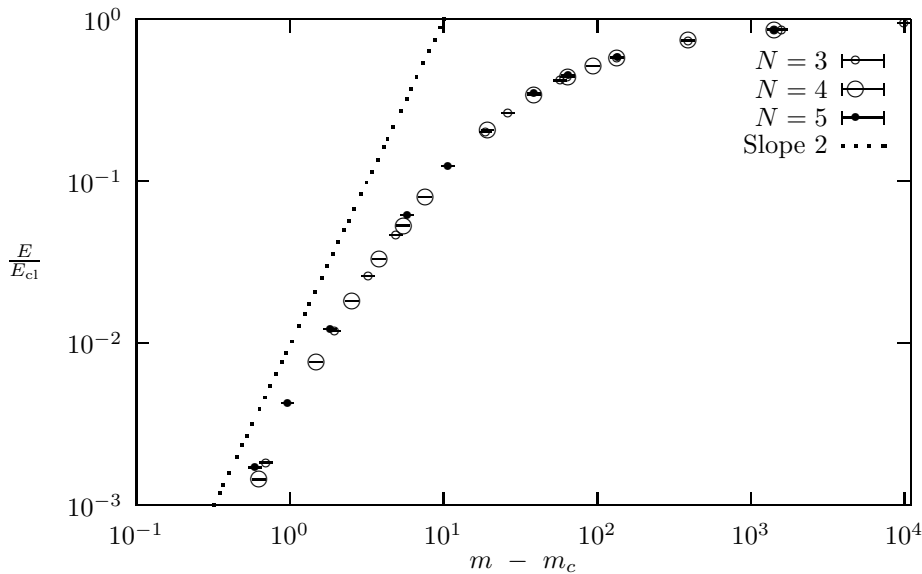

\centerline{\input ./log.e} 
\vskip 0.5cm
\caption[f1]{Double-logarithmic plot of $E_0$ for $N = 3,4$ and $5$ 
versus $m-m_{\rm c}$.}
\label{fig.log.e}
\end{figure}

On the other hand, in the $m \rightarrow \infty$ limit (vanishing de
Boer parameter), the harmonic approximation predicts
\begin{equation}
\label{eq.Harm.approx}
(E_0 - E_{\rm{cl}})~|_{m \rightarrow \infty} \propto m^{-1/2}.
\end{equation}
where $E_{\rm{cl}}$ is the classical ground state energy.  In other
words, on the basis of Eqs.~(\ref{eq.critical}) and
(\ref{eq.Harm.approx}) the cluster energy is expected to be linear in
the de Boer parameter in both the classical and extreme quantum
limits.  Indeed, figure~\ref{fig.comb} displays this remarkably dull
behavior for the square root of the normalized energy as a function of
the de Boer parameter over the whole range.  To display the critical
behavior of the energy in more detail, Figure \ref{fig.log.e} shows a
double logarithmic plot of $E$ versus $m - m_{\rm c}$.


Numerical values for average distance and the gyration radius
were obtained using the approximation
\begin{equation}
\langle r \rangle \approx 
2 \langle r \rangle_{\psi_{\rm T},0} - 
\langle r \rangle_{\psi_{\rm T},\psi_{\rm T}},
\label{eq.approxexp}
\end{equation}
where the first term on the right denotes the mixed expectation value
obtained by diffusion Monte Carlo, while the second term is the
variational expectation value of the cluster radius in the trial
state.

\begin{figure}
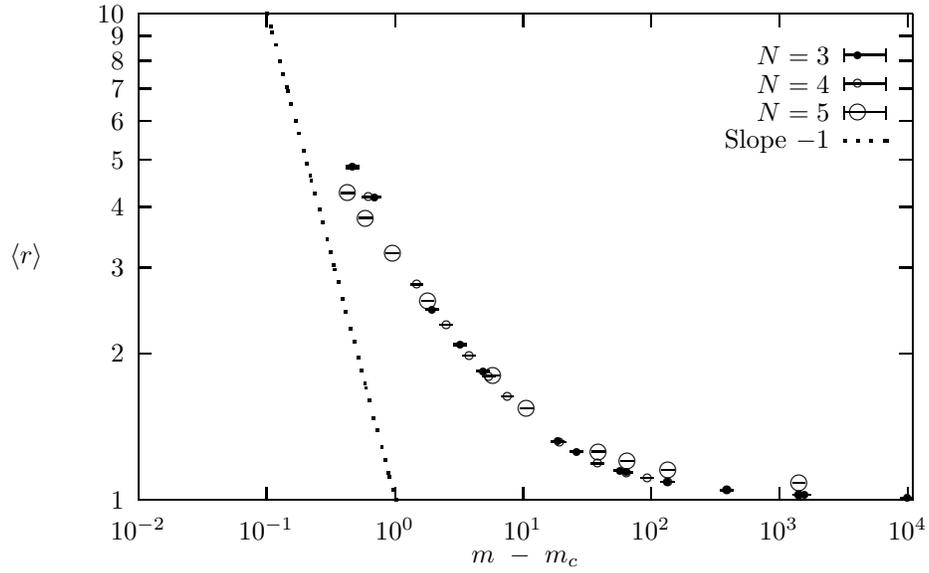

\centerline{\input ./log.r} 
\vskip 0.5cm
\caption[f1]{Double-logarithmic
plot of the approximate the average size of a cluster 
$\langle r \rangle$ defined as the average inter-atomic distance
versus $m-m_{\rm c}$ for $N = 3,4$ and $5$.}
\label{fig.log.r}
\end{figure}

\begin{figure}
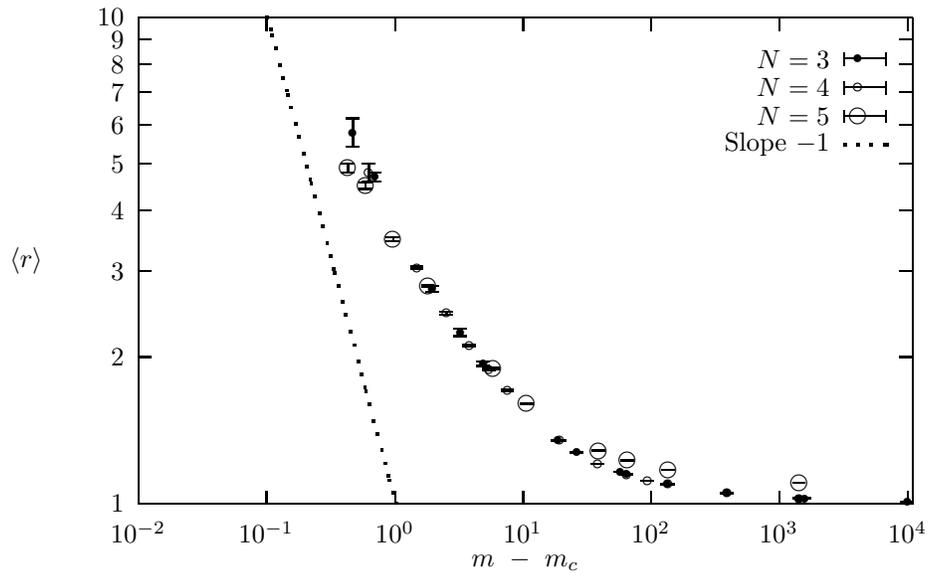

\centerline{\input ./log.R} 
\vskip 0.5cm
\caption[f1]{Double-logarithmic
plot of the approximate values of the average size of a cluster 
$\langle r \rangle$ defined as the gyration radius
versus $m-m_{\rm c}$ for $N = 3,4$ and $5$.}
\label{fig.log.R}
\end{figure}

Figs.~\ref{fig.log.r} and \ref{fig.log.R} are plots of the approximate
values of the average cluster size, as measured by the average
inter-particle distance and the gyration radius, versus inverse
$m-m_{\rm c}$.  The behavior displayed in the graphs is consistent with
the scaling law given in Eq.~(\ref{eq.critical}).  It should be noted
that there are apparent irregularities in the data points.  These can
be traced to irregularities in the quality of the wave functions, which
are a result of incomplete optimization. As far as the energy is
concerned, the quality of the trial wave functions only affects the
statistical accuracy of the estimates, but, as can be seen from
Eq.~\ref{eq.approxexp}, imperfections of the optimized trial wave
functions result in true errors of expectation values of quantities
that do not commute with the Hamiltonian, such as the size of the
clusters.

\acknowledgments

This research is supported by the National Science Foundation through
Grant \# DMR-9214669, by the Office of Naval Research.  It is a
pleasure to thank David Freeman, Alex Meyerovich and Cyrus Umrigar for
numerous valuable discussions.

\end{document}